\documentstyle[11pt,epsfig]{article}

\def\laq{~\raise 0.4ex\hbox{$<$}\kern -0.8em\lower 0.62ex\hbox{$\sim$}~}
\def\gaq{~\raise 0.4ex\hbox{$>$}\kern -0.7em\lower 0.62ex\hbox{$\sim$}~}

\def\beq{\begin{equation}}
\def\eeq{\end{equation}}
\def\bea{\begin{eqnarray}}
\def\eea{\end{eqnarray}}
\def\bean{\begin{eqnarray*}}
\def\eean{\end{eqnarray*}}

\def\l {\langle}
\def\re {\rangle}

\def \pa {\partial}

\def \Da {\Delta}

\def \ga {\gamma}

\def \Sg {\Sigma}

\def \noi {\noindent}

\def \Mq {{\cal M}_4}

    \def\be{\begin{equation}}
    \def\ee{\end{equation}}
    \def\ba{\begin{eqnarray}}
    \def\ea{\end{eqnarray}}

\newcommand{\eq}{\begin{equation}}
\newcommand{\eqx}{\end{equation}}
\newcommand{\eqn}{\begin{eqnarray}}
\newcommand{\eqnx}{\end{eqnarray}}

\newcommand{\Ups}{\Upsilon}

\newcommand{\rref}[1]{(\ref{#1})}

\newcommand{\lla}{\left\langle}
\newcommand{\rra}{\right\rangle}

\begin{document}

\begin{titlepage}

\begin{flushright}
BA-TH/641-11\\
CERN-PH-TH/2011-071\\
LPTENS-11/15
\end{flushright}

\vspace{0.5cm}

\begin{center}

\huge
 { Light-cone averaging in cosmology: \\
 formalism and applications}


\vspace{0.8cm}

\large{M. Gasperini$^{1,2}$, G. Marozzi$^{3}$, F. Nugier$^{4}$ and G. Veneziano$^{3,5}$}

\normalsize

\vspace{0.5cm}

{\sl $^1$Dipartimento di Fisica, Universit\`a di Bari, \\ 
Via G. Amendola 173, 70126 Bari, Italy}

\vspace{.1in}

{\sl $^2$Istituto Nazionale di Fisica Nucleare, Sezione di Bari, \\
Via G. Amendola 173, 70126 Bari, Italy}

\vspace{.1in}

{\sl $^3$ Coll\`ege de France, 11 Place M. Berthelot, 75005 Paris, France}

\vspace{.1in}

{\sl $^4$ Laboratoire de Physique Th\'eorique de l'\'Ecole Normale Sup\'erieure, 
CNRS UMR 8549,
24 Rue Lhomond, 75005 Paris, France}

\vspace{.1in}

{\sl $^5$CERN, Theory Unit, Physics Department, \\ CH-1211 Geneva 23, Switzerland}

\vspace*{1cm}


\begin{abstract}
\noi
We present a general gauge invariant formalism for defining cosmological averages  that are relevant for observations based on light-like signals. Such averages involve either null hypersurfaces corresponding to a family of past light-cones or compact surfaces given by their  intersection  with  timelike   hypersurfaces. Generalized Buchert-Ehlers commutation rules for  derivatives of these light-cone averages are given. After  introducing some adapted  ``geodesic light-cone" coordinates,  we  give explicit expressions for averaging the redshift to luminosity-distance relation and  the so-called ``redshift drift" in a generic inhomogeneous Universe.
\end{abstract}
\end{center}

\end{titlepage}

\newpage

\parskip 0.2cm


\section{Introduction}
\label{Sec1}
\setcounter{equation}{0}

It is by now well-known (see, for example, \cite{Ellis})  that averaging solutions of the full inhomogeneous  Einstein equations leads, in general, to different results from those obtained by solving the  averaged (i.e. homogeneous) Einstein equations. In particular,  the averaging procedure does not commute with the non-linear differential operators appearing in the Einstein equations and, as a result, the dynamics of the averaged geometry is affected by so-called ``backreaction'' terms, originating from the  inhomogeneities present
in the geometry and in the various components of the cosmic fluid.

Following the discovery of cosmic acceleration on large
scales, interest in  the possible effects of inhomogeneities for  interpreting the data themselves has considerably risen (see \cite{Review1,Review2} for some recent review papers). 
It has even been suggested (see e.g. \cite{3ref})  that the dynamical effects of the backreaction could replace, at least in part, the dark-energy sources as an explanation of  such a cosmic acceleration,
thereby  providing an elegant  solution to the well-known ``coincidence problem''.

As a consequence, much work has been done  in these last few years on trying to formulate a suitable ``averaged'' description of  inhomogeneous cosmology. In most of these works, following Buchert's seminal papers \cite{1}, the effective geometry emerging after the smoothing-out of local inhomogeneities has been determined by integrating over three-dimensional spacelike hypersurfaces. 

However, as pointed out long ago \cite{Maartens1,Maartens2}, a phenomenological reconstruction of the spacetime metric and of its dynamic evolution on a cosmological scale is necessarily based on past light-cone observations, since most of the relevant signals travel with the speed of light. Hence, the averaging  procedure should be possibly referred to a null hypersurface coinciding with a past light-cone rather than to some fixed-time spacelike hypersurface. Nonetheless, such a light-cone averaging procedure, whose importance has been repeatedly stressed in the specialized literature (see e.g. \cite{Review1,3}), has never been implemented in practice. 

The aim of this paper is to introduce a general (covariant and gauge invariant) prescription for averaging scalar objects on null hypersurfaces, to apply it to the past light-cone of a generic observer in the context of an inhomogeneous cosmological metric, and to provide the analog of the Buchert-Ehlers commutation rules \cite{4} for the derivatives of light-cone averaged quantities. We will also introduce an adapted system of coordinates (defining what  we call a  ``geodesic light-cone frame'', which can be seen as a particular specification of the ``observational coordinates'' introduced in \cite{Maartens1,Maartens3}) where the averaging prescription greatly simplifies, while keeping all the required degrees of freedom for applications to general inhomogeneous metric backgrounds.

In order to illustrate our light-cone averaging procedure we will propose, in particular, two possible physical applications: the averaging of the luminosity-redshift relation and that of the so-called 
``redshift drift'' parameter \cite{9} (see \cite{Uzan,QABCQ} for recent discussions), both evaluated for a generic inhomogeneous cosmological geometry. We will concentrate our attention on the realistic case in which the reference observer is geodesic, having in mind a model of light ray propagation based on the geometrical optics approximation (see e.g. \cite{KS,AppEt1933a}).

Finally, we stress that our intent, in the present paper, is not to give a detailed physical discussion and interpretation of the many phenomenological effects of averaging, and to bring arguments supporting (or disproving) plausible theoretical alternatives to the ``standard" dark-energy/cosmic acceleration scenario. We simply present a first step towards such an ambitious program, providing a formal procedure allowing an automatic  implementation of light-cone averaging, a procedure that -- to the best of our knowledge -- was missing in the present literature, and that we believe to be of some utility because of its covariance and model independence. A discussion of the possible phenomenological consequences of its application to specific cosmological scenarios will be presented in a forthcoming paper.

The paper is organized as follows. In Sect.\ref{Sec2} we recall the averaging prescription introduced in \cite{GMV1,GMV2} and we 
generalize it to different kinds of averaging on null hypersurfaces.
In Sect.\ref{Sec3} we give the covariant and gauge invariant version
of the Buchert-Ehlers commutation rules for light-cone averages. 
In Sect.\ref{Sec4} we introduce a coordinate system in which the light-cone averaging prescriptions and commutation rules take a much simpler form. In Sect.\ref{Sec5} we discuss a possible approach to the average of the luminosity-redshift relation and of the  redshift drift  on the light-cone of a geodesic observer. Our conclusive remarks are presented in Sect.\ref{Sec6}. 


\section{Gauge invariant light-cone averaging}
\label{Sec2}
\setcounter{equation}{0}

Let us first briefly recall  the approach given in \cite{GMV1,GMV2} to gauge invariant averaging on a three-dimensional spacelike hypersurface $\Sg(A)$, embedded in our four-dimensional spacetime 
$\Mq$. 
Assuming the hypersurface (or a spacelike foliation) to be defined by an equation involving a scalar field with timelike gradient $A(x)$:
\beq
A(x) - A_0 = 0,
\eeq
the gauge (and hypersurface-parametrization) invariant definition of  the integral of an arbitrary scalar $S(x)$ and of its average on such hypersurface was given in \cite{GMV1,GMV2}  as\footnote{In \cite{GMV2} the prescription introduced in \cite{GMV1} is used to give a covariant and gauge invariant generalization of the effective equations presented in \cite{1}. Such a generalization
has been recently used  to deal with  the backreaction of quantum fluctuations in an inflationary model \cite{FMVV}.}:
\begin{eqnarray}
&&
I(S;A_0) =  \int_{\Mq} d^4 x \sqrt{-g(x)}~\delta(A(x)-A_0)~\sqrt{-\partial_\mu A \partial^\mu A}~ S(x) ;
\nonumber \\ &&
\langle S \rangle_{A_0} = \frac{I(S; A_0)}{I(1; A_0)} 
\end{eqnarray}
(we are using the signature $(-,+,+,+)$). 
Here  the spatial hypersurface has no boundary. However, as shown in \cite{GMV1, GMV2}, a possible spatial boundary can be added through the following extension of the previous integral:
\beq
I(S;A_0; B_0) =  \int_{\Mq} d^4 x \sqrt{-g(x)} \delta(A(x)-A_0) \Theta(B_0- B(x)) \sqrt{-\partial_\mu A \partial^\mu A}
~S(x)\,,
\label{23}
\eeq
and similarly for the corresponding average ($\Theta$ is the Heaviside step function, and $B$ is a positive function of the coordinates, with spacelike gradient). As discussed in  \cite{GMV1,Marozzi}, this is still a gauge invariant expression if $B(x)$ transforms as a scalar,  while it gives violations of gauge invariance if $B$ is not a scalar and keeps the same form in different coordinate systems. Even in that case, however, gauge invariance violations go to zero  when we choose $B_0$ in such a way that the size of the spatial region goes to infinity \cite{GMV1,Marozzi}.

The above procedure unfortunately fails if $A(x) = A_0$ defines a null (lightlike) hypersurface, since in that case $\partial_\mu A \partial^\mu A =0$. In order to circumvent this problem let us start with a spacetime integral where the four-dimensional integration region is
bounded by two hypersurfaces, one  spacelike and the other one null (corresponding e.g. to the past light-cone of some observer). Let us choose, in particular, the region inside the past light-cone of the observer bounded in the past by the hypersurface $A(x)=A_0$: clearly   a gauge invariant definition of the integral of a scalar $S(x)$ over such a hypervolume can be written (in a useful notation generalizing the one used above) as 
\eq
I(S;-;A_0,V_0)= \int_{\Mq} d^4x \sqrt{-g}~ \Theta(V_0-V)  \Theta(A-A_0)  S(x),
\label{Intvol}
\eqx
where $V(x)$ is a (generalized advanced-time) scalar satisfying $ \partial_{\mu} V \partial^{\mu} V=0$, where $V_0$ specify the past light cone of a given observer, and where the ``$-$'' symbol on the l.h.s.  denotes the absence of delta-like window functions. 

\begin{figure}[t]
 \centering
\includegraphics[width=13cm]{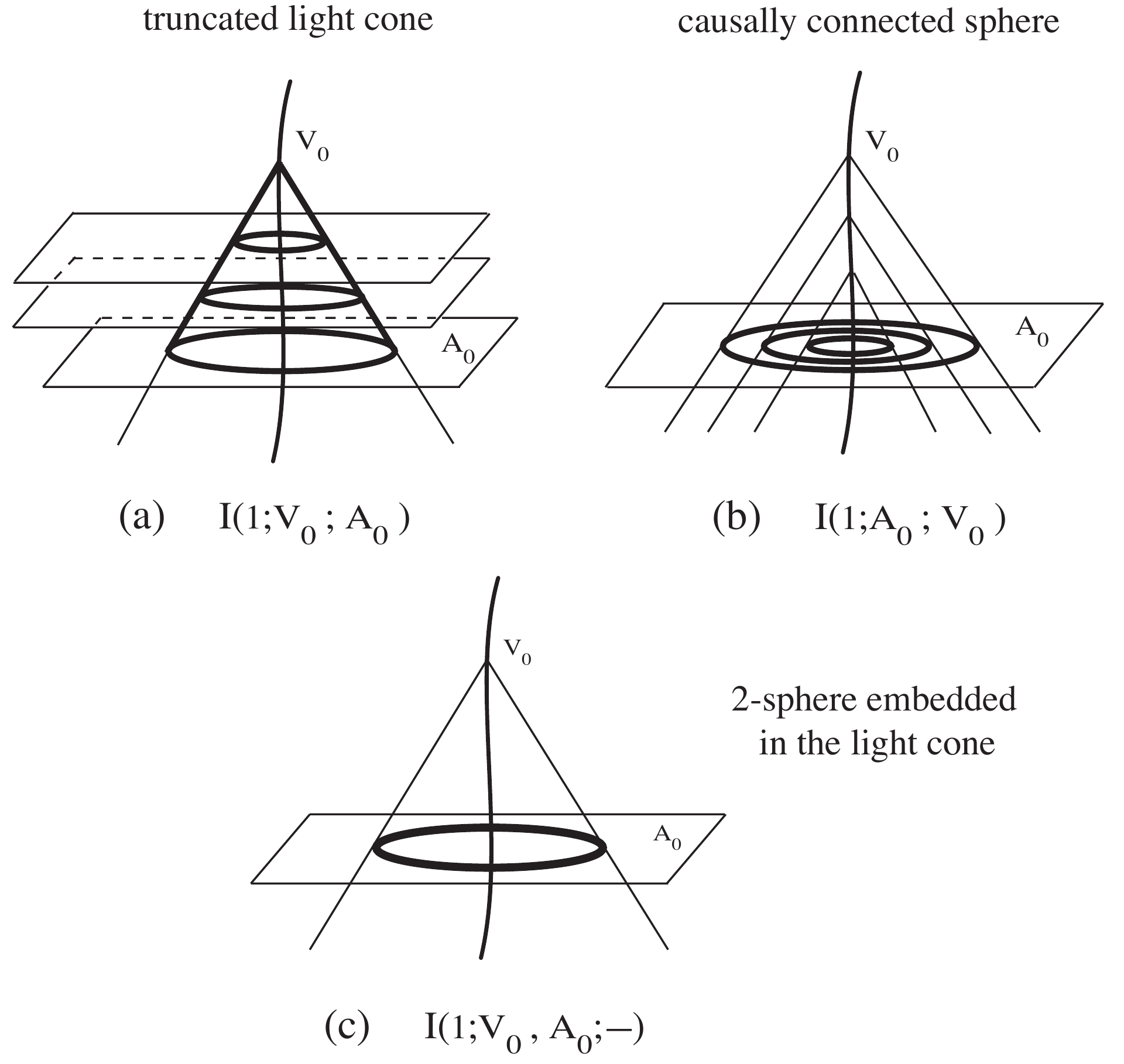}
 \caption{Three different light-cone averaging prescriptions.   
$(a)$: the average of Eq. (\ref{null3d}) on the light-cone itself starting from a given hypersurface in the past; $(b)$: the average of Eq. (\ref{null3dInv}) on the section of the hypersurface $A(x)=A_0$ which is causally connected with us; $(c)$: the average of Eq. (\ref{null2d}) on a 2-sphere embedded in the light-cone.}
 \label{fig1}
 \end{figure}

Starting with this hypervolume integral we can construct gauge invariant hypersurface and surface integrals by applying  to it appropriate differential operators -- or, equivalently,  by applying Gauss's theorem to the volume integral of a covariant divergence. An example of the latter, if we are interested into the variations of the volume averages along the flow lines normal to the reference hypersurface $\Sg(A)$, is obtained by replacing the scalar $S$ with the divergence of the unit normal to $\Sg$,
\beq
n_\mu =- {\pa_\mu A\over \sqrt{-\partial_\nu A \partial^\nu A} }, ~~~~~~~~~~~~~ n_\mu n^\mu=-1,
\label{vel}
\eeq
and leads to the identity:
\bea
&& \!\!\!\!\!\!\!\!\!\! \!\!\!\!\!\!\!\!\!\! \!\!\!\!\!\!\!\!\!\!
\int_{\Mq} d^4x \sqrt{-g}~ \Theta(V_0-V)  \Theta(A-A_0)  \nabla^{\mu}  n_{\mu} =
\nonumber \\ && ~~~~~~~~~~~~~~~~
= -\int_{\Mq} d^4x \sqrt{-g}\, \Theta(V_0-V) \delta(A-A_0)  
\sqrt{-\partial_\mu A \partial^\mu A} 
\nonumber \\ && ~~~~~~~~~~~~~~~~~~~~
  + \int_{\Mq} d^4x \sqrt{-g} \,\delta(V_0-V)  \Theta(A-A_0) 
\frac{-\partial_\mu V \partial^\mu A}{\sqrt{-\partial_\nu A \partial^\nu A}}.
\label{Gauss1}
\eea
Hence, if we start from Eq. (\ref{Intvol}), and we consider the variation of the average integral by shifting the light-cone 
$V=V_0$ along the flow lines defined by $n_\mu$,  we are led to define the hypersurface integral (with positive measure): 
\be
I(1;V_0;A_0)=\int_{\Mq} d^4x \sqrt{-g} \,\delta(V_0-V)  \Theta( A-A_0 )  
\frac{|\partial_\mu V \partial^\mu A|}{\sqrt{-\partial_\nu A \partial^\nu A}} \,.
\label{null3d}
\ee
Similarly, if we consider the variation of the average integral by shifting the hypersurface $A=A_0$ (along the same 
flow lines defined by $n_\mu$), we are led to another hypersurface integral: 
\be
I(1;A_0;V_0)=\int_{\Mq} d^4x \sqrt{-g}\, \Theta(V_0-V)  \delta( A-A_0 )  
\sqrt{-\partial_\mu A \partial^\mu A}.
\label{null3dInv}
\ee
In the first case, Eq. (\ref{null3d}), the integration region is on the light-cone itself, and it is spanned by the variation of  $\Sg(A_0)$ along its normal, at fixed light-cone $V_0$ (see Fig. \ref{fig1}, $(a)$). In the second case of Eq. (\ref{null3dInv}) -- which gives exactly the same integral as in Eq. (\ref{23}) with $V$ replacing $B$ --  the hypersurface $\Sg(A_0)$ is kept fixed, and the integration region describes the causally connected section of $\Sg$ spanned by the variation  of the light-cone hypersurface (see Fig. \ref{fig1}, $(b)$). 

Further differentiation also lead to the following invariant surface integral
\eq
I(1;V_0,A_0;-)=\int_{\Mq} d^4x \sqrt{-g}~ \delta (V_0-V)  \delta(A-A_0)  
|\partial_\mu V \partial^\mu A|\,, 
\label{null2d}
\eqx  
with a compact, two-dimensional integration region defined by the intersection of $\Sg(A_0)$ with the light-cone $V_0$ (Fig. \ref{fig1}, $(c)$). This integral, as well as the integrals of Eqs. (\ref{null3d}), (\ref{null3dInv}), is not only covariant and gauge invariant but also invariant under separate reparametrizations of the scalar fields
$A\rightarrow \tilde{A}(A)$ and $V\rightarrow \tilde{V}(V)$. Eq. (\ref{null2d}), in addition, is a particular case
of an invariant integration over an arbitrary codimension-2 hypersurface defined by the conditions $A^{(n)}(x) =  0$, $ n = 1,2$. In general, and in $D$ spacetime dimensions, such an integral  can be written as 
\eq
\int_{\cal{M}_D} d^Dx \sqrt{-g} \prod_n \delta( A^{(n)}(x) ) \sqrt{|\det \bar{g}^{pm}|}~;
~~~~~~~~
~ \bar{g}^{pm} \equiv  \partial_{\mu} A^{(p)} \partial_{\nu} A^{(m)} g^{\mu \nu} \; ,
\label{v2}
\eqx
(as can be shown by considering the induced metric on the $(D-2)$-hypersurface), and is invariant under the more general  reparametrizations 
$A^{(1)}\rightarrow \tilde{A}^{(1)}\left(A^{(1)},A^{(2)}\right)$ and 
$A^{(2)}\rightarrow \tilde{A}^{(2)}\left(A^{(1)},A^{(2)}\right)$. It can be easily checked that Eq. (\ref{v2}) reduces to (\ref{null2d}) if $D=4$ and if $A^{(1)} = A - A_0$ and $A^{(2)} = V- V_0$ are  scalar functions with timelike and null gradient, respectively. 

In order to make contact with Eqs. (\ref{null3d}), (\ref{null3dInv}),  it may be useful to remark that the integral (\ref{null2d})  can be also   obtained  starting from the  hypervolume integral (\ref{Intvol}) by considering the variation of the volume average along the flow lines normal to 
$\Sigma(A)$ for both $\Theta(A)$ and $\Theta(V)$, namely by using the following window function:
\begin{eqnarray}
&& \!\!\!\!\!\!\!\!\!\! \!\!\!\!\!\!\!\!\!\! 
 -n^{\mu} \nabla_\mu \Theta(A(x)-A_0)n^{\mu} \nabla_\mu \Theta(V_0-V(x))= 
\nonumber \\
&&   ~~~~~~~~~~~~~~ =
\sqrt{-\partial_\mu A \partial^\mu A} \, \delta(A(x)-A_0) \frac{-\partial_\mu V 
\partial^\mu A}{\sqrt{-\partial_\mu A \partial^\mu A}}  \delta(V_0-V(x)) \,. 
\end{eqnarray}

We note, finally, that averages of a scalar $S$ over different (hyper)surfaces are trivially defined -- with self explanatory notation -- by:
\bea
\langle S \rangle_{V_0,A_0} &=& \frac{I(S;V_0,A_0;-)}{I(1;V_0,A_0;-)}~;~
\label{211} \\  
\langle S \rangle_{V_0}^{A_0} &=& \frac{I(S;V_0;A_0)}{I(1;V_0;A_0)}~;~
\label{212}\\ 
\langle S \rangle_{A_0}^{V_0} &=& \frac{I(S;A_0;V_0)}{I(1;A_0;V_0)}.
\label{avnull2d}
\label{213}
\eea


\section{Buchert-Ehlers commutation rules on the light-cone}
\label{Sec3}
\setcounter{equation}{0}

For the phenomenological applications of the following sections we are interested, in particular, in the derivatives of the averages defined in 
(\ref{211}) and (\ref{212}). To this purpose, let us first consider  the derivatives of $I(S;V_0,A_0;-)$ and $I(S;V_0;A_0)$  with respect to  $A_0$ and $V_0$ (quantities that, like the starting integrals themselves, are covariant and gauge invariant).

We will use the identities (here $k_{\mu} \equiv \partial_{\mu} V$):
\begin{equation}
\label{RelDeltaThetaPrime}
\delta'(A-A_0) = \frac{k^{\mu}\partial_{\mu}\delta(A-A_0)}{k^{\nu}\partial_{\nu}A} 
\,\,\,\,\,\,,\,\,\,\,\,\,
\delta(A-A_0) = \frac{k^{\mu}\partial_{\mu}\Theta(A-A_0)}{k^{\nu}\partial_{\nu}A} \,,
\end{equation}
\be 
 \delta'(V-V_0) = \frac{\partial^{\mu} A \partial_{\mu}\delta(V-V_0)}{k^{\nu}\partial_{\nu}A}
 \,\,\,\,\,\,,\,\,\,\,\,\,
 \delta(V-V_0) = \frac{\partial^{\mu} A \partial_{\mu}\Theta(V-V_0)}{k^{\nu}\partial_{\nu}A}\,,
\label{RelDeltaThetaPrimeV}
\ee
and the relation $k^{\mu}\partial_{\mu}\delta(V-V_0) = k^{\mu}k_{\mu}\delta'(V-V_0) = 0$. Integrating by parts, we then obtain
\begin{eqnarray}
& &\frac{\partial}{\partial A_0} I(S;V_0,A_0;-) = 
I\left(\frac{k \cdot \partial S}{k \cdot \partial A};V_0,A_0;-\right)
+I\left(\frac{\nabla \cdot k}{k \cdot \partial A} S;V_0,A_0;-\right) \,,\nonumber \\ 
& &\frac{\partial}{\partial A_0} I(S;V_0;A_0) = 
I\left(\frac{k \cdot \partial S}{k \cdot \partial A};
V_0;A_0\right)
+I\left(\left[\nabla \cdot k -\frac{1}{2} 
\frac{k^\mu \partial_\mu \left((\partial A)^2\right)}{(\partial A)^2} 
\right]\frac{S}{k \cdot \partial A};V_0;A_0\right),
\nonumber \\ & &
\end{eqnarray} 
where, to simplify notations, we have introduced the following definitions: 
$k^\mu \partial_\mu S=k \cdot \partial S$, 
$k^\mu \partial_\mu A=k \cdot \partial A$, 
$\partial_\mu A \partial^\mu S=\partial A \cdot \partial S$, 
$\partial_\mu A \partial^\mu A=(\partial A)^2$ , 
$\nabla_\mu k^\mu = \nabla \cdot k$ 
and $\Box=\nabla^\mu \nabla_\mu$.

Using these results we can write a generalized version of the Buchert-Ehlers commutation rule \cite{4}  for light-cone average. In the case of averages defined over the integration domain (\ref{null2d}) we find 
\begin{equation}
\label{LCeq1}
\frac{\partial}{\partial A_0} \langle S \rangle_{V_0,A_0} = \left\langle \frac{k \cdot \partial S}{k \cdot \partial A}\right\rangle_{V_0,A_0} + \left\langle \frac{\nabla \cdot k}{k \cdot \partial A} S \right\rangle_{V_0,A_0} - \left\langle \frac{\nabla \cdot k}{k \cdot \partial A} \right\rangle_{V_0,A_0}
\langle S \rangle_{V_0,A_0},
\end{equation}
 while for 3-dimensional averages over the domain (\ref{null3d}) we obtain:
 \begin{eqnarray}
\frac{\partial}{\partial A_0} \langle S \rangle_{V_0}^{A_0} &=& \left\langle \frac{k \cdot \partial S}{k \cdot \partial A}\right\rangle_{V_0}^{A_0} + \left\langle \left[ \nabla \cdot k
- \frac{1}{2}\frac{k^{\mu}\partial_{\mu}\left((\partial A)^2\right)}{(\partial A)^2}\right]
\frac{S}{k \cdot \partial A} \right\rangle_{V_0}^{A_0} 
\nonumber \\
& &  - \left\langle \left[ \nabla \cdot k
- \frac{1}{2}\frac{k^{\mu}\partial_{\mu}\left((\partial A)^2\right)}{(\partial A)^2}\right]
\frac{1}{k \cdot \partial A} \right\rangle_{V_0}^{A_0}
\left\langle S \right\rangle_{V_0}^{A_0}.
\label{LCeq2}
\end{eqnarray}

Following a similar procedure we can also evaluate the generalization of the Buchert-Ehlers commutation rule for the derivative of $I(S;V_0,A_0;-)$ and $I(S;V_0;A_0)$  with respect to  $V_0$.
As we shall see in Sect. \ref{Sec5}, such derivatives, together with the previous ones, may be useful in evaluating an averaged version of the  ``redshift drift" parameter.
After some calculations we are led to 
\begin{eqnarray}
\frac{\partial}{\partial V_0} \langle S \rangle_{V_0,A_0} &=& 
\left\langle \frac{\partial A \cdot \partial S}{k \cdot \partial A} \right\rangle_{V_0,A_0} 
-\lla k \cdot \partial S \frac{(\partial A)^2}{(k \cdot \partial A)^2} \rra_{V_0,A_0}
\nonumber \\ & & 
+\left\langle
\left[\Box A-\nabla_\mu \left(
k^\mu  
\frac{(\partial A)^2}{k \cdot \partial A}\right)\right] 
\frac{S}{k \cdot \partial A}
\right\rangle_{V_0,A_0}
\nonumber \\ & & 
-\left\langle
\left[\Box A-\nabla_\mu \left(
k^\mu  
\frac{(\partial A)^2}{k \cdot \partial A}\right)\right] 
\frac{1}{k \cdot \partial A}
\right\rangle_{V_0,A_0}
\langle S \rangle_{V_0,A_0} \,,
\label{LCeq3}
\end{eqnarray}
and
\begin{eqnarray}
\label{LCeq4}
\frac{\partial}{\partial V_0} \lla S \rra_{V_0}^{A_0} & = & 
\left\langle \frac{\partial A \cdot \partial S}{k \cdot \partial A} \right\rangle_{V_0}^{A_0} 
-\lla k \cdot \partial S \frac{(\partial A)^2}{(k \cdot \partial A)^2} \rra_{V_0}^{A_0}
\nonumber \\ & & 
+\left\langle
\left[\Box A-\partial^\mu A \partial_\mu \ln\left((\partial A)^2\right)
\right] \frac{S}{k \cdot \partial A}
\right\rangle_{V_0}^{A_0}
\nonumber \\
& &
- \left\langle
\left[\nabla \cdot k \frac{(\partial A)^2}{(k \cdot \partial A)^2}+\frac{1}{2} k^\mu \partial_\mu 
\left(\frac{(\partial A)^2}{(k \cdot \partial A)^2}\right)\right] S \right\rangle_{V_0}^{A_0}
\nonumber \\
& &  
- \left\langle
\left[\Box A-\partial^\mu A \partial_\mu \ln\left((\partial A)^2\right)
\right] \frac{1}{k \cdot \partial A}
\right\rangle_{V_0}^{A_0}  \left\langle S \right\rangle_{V_0}^{A_0} 
\nonumber \\
& & 
+ \left\langle
\left[\nabla \cdot k \frac{(\partial A)^2}{(k \cdot \partial A)^2}+\frac{1}{2} k^\mu \partial_\mu 
\left(\frac{(\partial A)^2}{(k \cdot \partial A)^2}\right)\right] \right\rangle_{V_0}^{A_0}
 \left\langle S \right\rangle_{V_0}^{A_0}\; .
\end{eqnarray}
In a similar way one could also derive equations for the derivatives of the average $ \lla S \rra_{A_0}^{V_0}$. However, since we will not discuss  applications of those, we will not give  their explicit form here.

 
\section{Geodesic light-cone coordinates}
\label{Sec4}
\setcounter{equation}{0}

\subsection{Definition of geodesic light-cone gauge}

We now turn to a special (adapted) coordinate system, $x^\mu= (w,\tau,\theta^a)$, $a=1,2$,  in which the previous equations take a  simpler form.  In this sense they will play a role similar to the one played by synchronous gauge coordinates for spatial averages with respect to geodesic observers \cite{Marozzi}. We are interested in coordinates such that the level sets of one of them define the past light-cones, while those of another coordinate define a set of geodesic observers. We claim that such coordinates, that we will call geodesic light-cone (GLC) coordinates, are defined by the metric:
\eq
\label{LCmetric}
d s^2 = \Ups^2 d w^2 - 2 \Ups d w d \tau + \gamma_{ab}(d \theta^a - U^a d w)(d \theta^b - U^b d w)~;~~~~~~~~~ a, b = 1,2 \; .
\label{41}
\eqx
This metric depends on� six arbitrary functions ($\Ups$, the two-dimensional vector $U^a$� and the symmetric tensor $\gamma_{ab}$) and corresponds to a complete gauge� fixing (modulo residual transformations involving non-generic functions of all the coordinates) of the so-called observational coordinates\footnote{Note that our coordinates $\theta^a$ are not necessarily ``observational'', in general,� but they can be reduced to such form (e.g. to standard spherical coordinates, parallelly propagated along the observer world-line) through an appropriate relabelling of null generators \cite{Maartens3}.}  discussed in detail in� \cite{Maartens1,Maartens3}. In matrix form,  the metric and its inverse are given by:
\eq
g_{\mu\nu} =
\left(
\begin{array}{ccc}
\Ups^2 + U^2 & - \Ups & -U_b \\
-\Ups & 0 & \vec{0} \\
-U_a & \vec{0} & \gamma_{ab} \\
\end{array}
\right),~~ ~~~~~
g^{\mu\nu} =
\left(
\begin{array}{ccc}
0 & -1/\Ups & \vec{0} \\
-1/\Ups & -1 & -U^b/\Ups \\
\vec{0} & -U^a/\Ups & \gamma^{ab} \\
\end{array}
\right) ,
\eqx
where $\gamma^{ab}$ is the inverse of $\gamma_{ab}$. Clearly $w$ is a null coordinate (i.e.  $\pa_\mu w \pa^\mu w=0$). More interestingly, one can check that $\partial_{\mu} \tau$ defines a geodesic flow, i.e. that
\beq
\left( \pa^\nu \tau\right) \nabla_\nu \left( \pa_\mu \tau\right) \equiv 0, 
\eeq
as a consequence of $g^{\tau \tau} = -1$. Thus an observer defined by constant $\tau$ spacelike hypersurfaces is in geodesic motion. Also note that  $\sqrt{-g} = \Ups \sqrt{|\gamma|}$, with $\gamma=\det\gamma_{ab}$.

In order to understand the geometric meaning of these variables, it is useful to  consider the limiting case  of  a spatially flat Friedmann-Lema\^itre-Robertson-Walker (FLRW) Universe, in the cosmic time gauge, with scale factor $a(t)$. 
Such a limit is easily reproduced by Eq. (\ref{41}) by setting 
\bea
&&
w= r+\eta, ~~~~~~~~~~~\tau=t, ~~~~~~~~~~~ \Ups = a(t), ~~~~~~~~~~~ U^a=0,
\nonumber \\ &&
\gamma_{ab} d \theta^a d\theta^b = a^2(t) r^2 (d \theta^2 +\sin^2 \theta d\phi^2),
\label{FR}
\eea
where $\eta$ is the conformal time of the homogeneous metric: $d\eta= dt/a$. 

In these coordinates a past light-cone hypersurface is specified by the condition $w=w_0$, and the momentum of a photon traveling on it,  being proportional to $k_\mu=\partial_{\mu}w$, is orthogonal both to itself ($k_{\mu}k^{\mu}=0$) and to the 2-surface generated by $\partial_{\theta^a}$  ($\partial_{\mu} {\theta^a} k^{\mu}=0$).
The velocity of a  generic observer defined by a scalar $A$ as in Eq. (\ref{vel}) satisfies:
\beq
k_{\mu}n^{\mu} = \frac{1}{\Ups} \frac{\partial_{\tau} A}{\sqrt{-(\partial A)^2}}  \,.
\eeq
For a  geodesic observer  $A$ depends only on $\tau$ and we can always set $A=\tau$, so the above relation simplifies to:
\be
k_{\mu}n^{\mu} = \Ups^{-1} ~. 
\label{45}
\ee
These equations will be used in Sect. \ref{Sec5} in connection with the redshift and luminosity distance. We also note for later use that, in these coordinates, the covariant divergence  of $k^{\mu}$  takes the simple form:
\eqn
\nabla_{\mu} k^{\mu}
=  -\frac{1}{\Ups } \partial_{\tau}\left(\ln \sqrt{|\gamma|} \right)\,.
\eqnx


\subsection{Average equations in the GLC gauge}
\label{Sec41}

In the GLC gauge the averaging integrals introduced in Sect. \ref{Sec2} greatly simplify, especially in the case where the reference hypersurface $\Sg(A)$ defines a geodesic observer. We will concentrate on the integrals  (\ref{null2d}) and (\ref{null3d}) which, setting $V= w, V_0=w_0$, are now given by: 
\begin{eqnarray}
& &\!\!\!\!\!\!\!\!\!\!\!\!\!\!\!\!\!\!I(S;w_0,A_0;-)=\int d^2\theta  dw d\tau \sqrt{|\gamma|}~
\left|\partial_\tau A \right|
 \delta (w-w_0)  \delta( A-A_0 ) S  ~ , 
\label{null2d_SLC} \\
& &\!\!\!\!\!\!\!\!\!\!\!\!\!\!\!\!\!\!I(S;w_0;A_0)=\int d^2\theta  dw d\tau \sqrt{|\gamma|}~
\frac{|\partial_\tau A|}{\sqrt{-(\partial A)^2}}
 \delta (w-w_0)  \Theta( A-A_0 ) S. 
\label{null3d_SLC}
\end{eqnarray}
For a geodesic reference observer, with $A=\tau$ (and $A_0=\tau_0$), we obtain:
 \begin{eqnarray}
I(S;w_0,\tau_0;-)&=&\int d^2\theta  \sqrt{|\gamma(w_0, \tau_0, \theta^a)|}~
 S(w_0, \tau_0, \theta^a) , 
\label{null2d_SLC_geodetic} \\
I(S;w_0;\tau_0)&=&\int d^2\theta d\tau \sqrt{|\gamma(w_0, \tau, \theta^a)|}~
\Theta( \tau-\tau_0 ) S(w_0, \tau, \theta^a).
\label{null3d_SLC_geodetic}
\end{eqnarray}
The generalized  Buchert-Elhers commutation rules introduced in Sect. \ref{Sec3}  also simplify as illustrated below.


\paragraph{Average on the 2-sphere embedded in the light-cone}\

Let us consider separately, as before, derivatives with respect to $A_0$ and to $V_0= w_0$. Starting with $A_0$, Eq.  \rref{LCeq1} can now be written as
\eq
\partial_{A_0} \lla S \rra_{w_0,A_0} = \lla \frac{\partial_{\tau} S}{\partial_{\tau} A}\rra_{w_0,A_0} + \lla S \frac{\partial_{\tau}\ln\sqrt{|\gamma|}}{\partial_{\tau} A}\rra_{w_0,A_0} - \lla \frac{\partial_{\tau}\ln\sqrt{|\gamma|}}{\partial_{\tau} A}\rra_{w_0,A_0} \lla S \rra_{w_0,A_0},
\eqx
and for the special case 
where $\Sigma(A)$ defines a geodesic observer (with $A=\tau$ and $A_0=\tau_0$), it reduces to
\eq
\partial_{\tau_0} \lla S \rra_{w_0,\tau_0} = \lla \partial_{\tau} S \rra_{w_0,\tau_0} + \lla S \partial_{\tau}\ln\sqrt{|\gamma|}\rra_{w_0,\tau_0} - \lla \partial_{\tau}\ln\sqrt{|\gamma|}
\rra_{w_0,\tau_0}
 \lla S \rra_{w_0,\tau_0} \,.
 \label{413}
\eqx
Note that, in this last case with $A=\tau$,  the partial derivative with respect to $A_0$ reduces to a standard derivative with respect to the proper (cosmic) time if we consider the limit of a homogeneous FLRW Universe (see Eq. (\ref{FR})).

In the geodesic case we have also the following simplification for  the derivative with respect to $w_0$ as given in Eq. (\ref{LCeq3}):
\begin{eqnarray}
\partial_{w_0} \lla S \rra_{w_0,\tau_0} &=& \lla \partial_{w} S + U^a \partial_a S \rra_{w_0,\tau_0} + \lla S \partial_{w}\ln\sqrt{|\gamma|}\rra_{w_0,\tau_0} 
\nonumber \\ & &
+ \lla S \left[ \partial_a U^a +U^a \partial_a \ln\sqrt{|\gamma|}\right] \rra_{w_0,\tau_0} 
- \lla \partial_{w}\ln\sqrt{|\gamma|}\rra_{w_0,\tau_0} 
 \lla S \rra_{w_0,\tau_0}
\nonumber \\ & & 
- \lla \partial_a U^a +U^a \partial_a \ln\sqrt{|\gamma|} \rra_{w_0,\tau_0}
 \lla S \rra_{w_0,\tau_0} \,.
 \label{414}
\end{eqnarray}


\paragraph{Average on the truncated light-cone}\

For a generic hypersurface we cannot specify $(\pa A)^2$, and Eq.  
(\ref{LCeq2}) just simplifies as
\eqn
\partial_{A_0} \lla S \rra_{w_0}^{A_0} & = & \lla \frac{\partial_{\tau} S}{\partial_{\tau} A}\rra_{w_0}^{A_0} + \lla S \frac{\partial_{\tau}\ln\sqrt{|\gamma|}- \partial_{w}(\partial A \cdot \partial A)}{\partial_{\tau} A}\rra_{w_0}^{A_0} \nonumber \\
& & - \lla \frac{\partial_{\tau}\ln\sqrt{|\gamma|}- \partial_{w}(\partial A \cdot \partial A)}{\partial_{\tau} A}\rra_{w_0}^{A_0} \lla S \rra_{w_0}^{A_0}.
\eqnx
However, in the case 
where $\Sigma(A)$ defines a geodesic observer (with $A=\tau$ and $A_0=\tau_0$) one obtains:
\eq
\partial_{\tau_0} \lla S \rra_{w_0}^{\tau_0} = \lla \partial_{\tau} S \rra_{w_0}^{\tau_0} + \lla S \partial_{\tau}\ln\sqrt{|\gamma|} \rra_{w_0}^{\tau_0} - \lla \partial_{\tau}\ln\sqrt{|\gamma|} \rra_{w_0}^{\tau_0} \lla S \rra_{w_0}^{\tau_0}.
\eqx
Comparing with Eq. (\ref{413}) we find,  for 
this last case, the same commutation rule independently of the averaging prescription used.

Finally, for a hypersurface $\Sigma(A)$ associated to a 
geodesic observer, we  also have the following simplification of Eq. (\ref{LCeq4}):
\begin{eqnarray}
\partial_{w_0} \lla S \rra_{w_0}^{\tau_0} &=& \lla \partial_{w} S + U^a \partial_a S \rra_{w_0}^{\tau_0} + \lla S \partial_{w}\ln\sqrt{|\gamma|}\rra_{w_0}^{\tau_0} 
\nonumber \\ & &
+ \lla S \left[ \partial_a U^a +U^a \partial_a \ln\sqrt{|\gamma|}\right] \rra_{w_0}^{\tau_0} 
-  \lla \partial_{w}\ln\sqrt{|\gamma|}\rra_{w_0}^{\tau_0} 
 \lla S \rra_{w_0}^{\tau_0}
\nonumber \\ & & 
- \lla \partial_a U^a +U^a \partial_a \ln\sqrt{|\gamma|} \rra_{w_0}^{\tau_0}
 \lla S \rra_{w_0}^{\tau_0} \,.
 \label{416}
\end{eqnarray}
Hence, we obtain the same commutation rule (see Eq. (\ref{414})) even for derivatives with respect to $w_0$.


\section{Some physical applications}
\label{Sec5}
\setcounter{equation}{0}

Information about the large scale structure of our Universe reaches us travelling along the null geodesics of a possibly inhomogeneous spacetime. Hence, an averaged description of the cosmological geometry should unavoidably make use of some light-cone averaging procedure, like those illustrated in the previous sections. In this section we will suggest two possible applications of the previous formalism: we will consider, in particular, the averaging of the luminosity-distance redshift relation  and the averaging of the redshift drift parameter for a generic inhomogeneous  cosmology. Note that in this section we will be dealing with  the $\tau$ coordinate both at the source and at the observer's position. They play the same role as $A_0 \equiv \tau_0$ of the previous sections, and will be denoted by $\tau_s$ and $\tau_0$ respectively.


\subsection{The redshift to luminosity-distance relation}

It is well known that the scalar redshift parameter $z$, for a photon with momentum $k_\mu$ emitted by a source $S$ and received by the observer $O$, depends on the scalar product between $k_\mu$ and the four-velocity $n_\mu$ of a local reference observer, and is given in general (with self-explanatory notation) by:
\eq
1+z = \frac{\left(k_\mu n^\mu\right)_s}{\left(k_\nu n^\nu  \right)_0}.
\label{redshift}
\eeq
Note that this object  may be regarded as a bi-scalar, as it depends on the ratio of the same scalar quantity calculated in two different spacetime points. In the GLC gauge of Sect. \ref{Sec4}, and in the physically interesting case of a geodesic observer (that we will consider hereafter), we can use   Eq. (\ref{45}) and we thus obtain: 
\eq
1+z = \frac{\Ups_0}{\Ups_s}\,,
\label{zratioY}
\eqx
where $\Ups_0=\Ups(w_0,\tau_0,\theta^a_0)$ and $\Ups_s=\Ups(w_0,\tau_s,\theta^a_s)$.

In a cosmological context we can define at least three different distances of physical interest (see e.g. \cite{Weinberg}): $d_s$, namely  the angular distance of the source as seen from the observer (in the FLRW limit of Eq. (\ref{FR}) we have $d_s = r a(t_s)$, where  $r$ is the comoving radial distance of the source); $d_o$, namely  the angular distance of the observer ``as seen from'' the source (in the FLRW limit  we have $d_o = r a(t_0)$); and $d_L$, the so-called luminosity distance. These three observational distances are always related to each other, independently of the  given cosmological model,  by redshift factors \cite{AppEt1933a,AppEt1933b}:
\eq
d_L = (1+z) d_o = (1+z)^2 d_s,
\eqx 
as a consequence of the Etherington  reciprocity law \cite{Et1933}.

On the other hand, in a general spacetime, we can also identify (point-by-point on the 2-sphere) the square of the angular distance of the source with the ratio of the intrinsic cross-sectional area element to the subtended solid angle element,  according to $d_s^2={d {\cal A}}/{d \Omega}$ (see e.g.  \cite{KS}). In our GLC gauge  we have $k^\mu=(0, -1/\Upsilon, 0, 0)$, and, as already mentioned, the 2-sphere embedded in the light-cone (corresponding to the integration region of Eq. (\ref{null2d})) is orthogonal to the photon momentum, so that
the cross-sectional area element is proportional to $\sqrt{|\gamma|} d \theta^1  d \theta^2$. Therefore, $d_s^2$ can be written as 
\be
d_s^2(w=w_0, \tau_s, \theta^a)=\lim_{\rho \rightarrow 0} \, \rho^2 
\frac{\sqrt{|\gamma_s|}}{\sqrt{|\gamma (\rho)|}},
\label{d_s2}
\ee
where $\rho$ is the proper radius of an infinitesimal sphere centered around the observer, $w=w_0$ defines the past light-cone connecting source and observer, and $\tau=\tau_s$ defines the spacelike hypersurface normal to $n^\mu$ at the source position. Eq. (\ref{d_s2}) easily reduces to $d_s = r a(t_s)$ in the FLRW case of Eq. (\ref{FR}).

It is appropriate to mention that in the present discussion we are neglecting the possible occurrence of caustics in the past light cone of our observer, which would affect the assumed relation between area distance and angular size distance. The possible relevance of this interesting effect, and its dependence on gravitational lensing and on the distribution of inhomogeneities has been addressed in detail in \cite{caustic}.

The relevant variables related to supernovae observations are the luminosity distance $d_L$ and the redshift $z$. 
Their relation has been studied within a gauge invariant approach, for a linearly perturbed FLRW metric, in \cite{Sasaki}. Averaging their values on the two-sphere embedded in the light-cone, and using the above results, we obtain, for a general inhomogeneous metric background,
\be
\langle d_L \rangle_{w_0,\tau_s}=\frac{\int d^2\theta  \sqrt{|\gamma(w_0, \tau_s, \theta^a)|}
\left[ \Ups^{2}(w_0, \tau_0, \theta^a)/ \Ups^{2}(w_0, \tau_s, \theta^a)\right]
d_s(w_0, \tau_s, \theta^a)}{\int d^2\theta \sqrt{|\gamma(w_0, \tau_s, \theta^a)|}},
\label{Int_dL} 
 \ee
where $d_s$ is given by Eq.(\ref{d_s2}), and
 \be
 \langle 1+ z \rangle_{w_0,\tau_s}=\frac{\int d^2\theta \sqrt{|\gamma(w_0, \tau_s, \theta^a)|}~
 \left[ \Ups(w_0, \tau_0, \theta^a)/ \Ups(w_0, \tau_s, \theta^a)\right]}
{\int  d^2\theta \sqrt{|\gamma(w_0, \tau_s, \theta^a)|}}.
\label{Int_z} 
 \ee
Since our coordinates should not be pathological near $w=w_0$ and $\tau=\tau_0$ we should have $\lim_{\tau\rightarrow \tau_0} 
\Ups (w=w_0, \tau, \theta^a)=\Ups (w=w_0, \tau_0)\equiv\Ups_0$ indepent of $\theta^a$. Hence, 
 in Eq. (\ref{Int_dL}) and (\ref{Int_z}), the factor $\Upsilon(w_0, \tau_0, \theta^a)=\Upsilon_0$ behaves like a constant,  
leaving a well defined integral of a scalar object over  the 2-sphere.

The above expressions can be used to see how the usual redshift to luminosity-distance relation gets affected by the inhomogeneities of the cosmological geometry. To this purpose it is also possible (and probably more convenient) to consider averages of the luminosity distance on constant redshift surfaces\footnote{We would like to thank  Misao Sasaki for stressing this point with us.}. 
In that case the geodesic observers measuring the redshift do not coincide any longer with the ``observers" associated to the flow lines of the  reference hypersurface $\Sigma(A)$, chosen to specify the averaging region. Even in such a case, however, our general formalism provides a rather simple averaging prescription by identifying the scalar $A$ in Eq. (\ref{211}) with $k_\mu u^\mu$ (where now $u^\mu$ is the velocity of a geodesic observer), and the quantity to be averaged with $d_L$ (or with some convenient observable related to it).

The straightforward calculation simplifies considerably in our gauge giving, for instance,
\be
\langle d_L \rangle_{w_0,z}= (1+z)^2 \frac{\int d^2\theta  \sqrt{|\gamma(w_0, \tau (z, w_0, \theta^a), \theta^a) |}
d_s(w_0,  \tau (z, w_0, \theta^a), \theta^a)}{\int d^2\theta  \sqrt{|\gamma(w_0, \tau (z, w_0, \theta^a), \theta^a )|}}
\label{Int_dLz} \, ,
 \ee
where $\tau (z, w_0, \theta^a)$ is the solution of:
\be
 \frac{\Ups(w_0, \tau , \theta^a)}{\Ups_0} = \frac{1}{1+z} \, .
\ee
We plan to come back to an explicit application of our formalism to this problem in the near future.
 
\subsection {Redshift drift}

\begin{figure}[ht]
 \centering
\includegraphics[width=12cm]{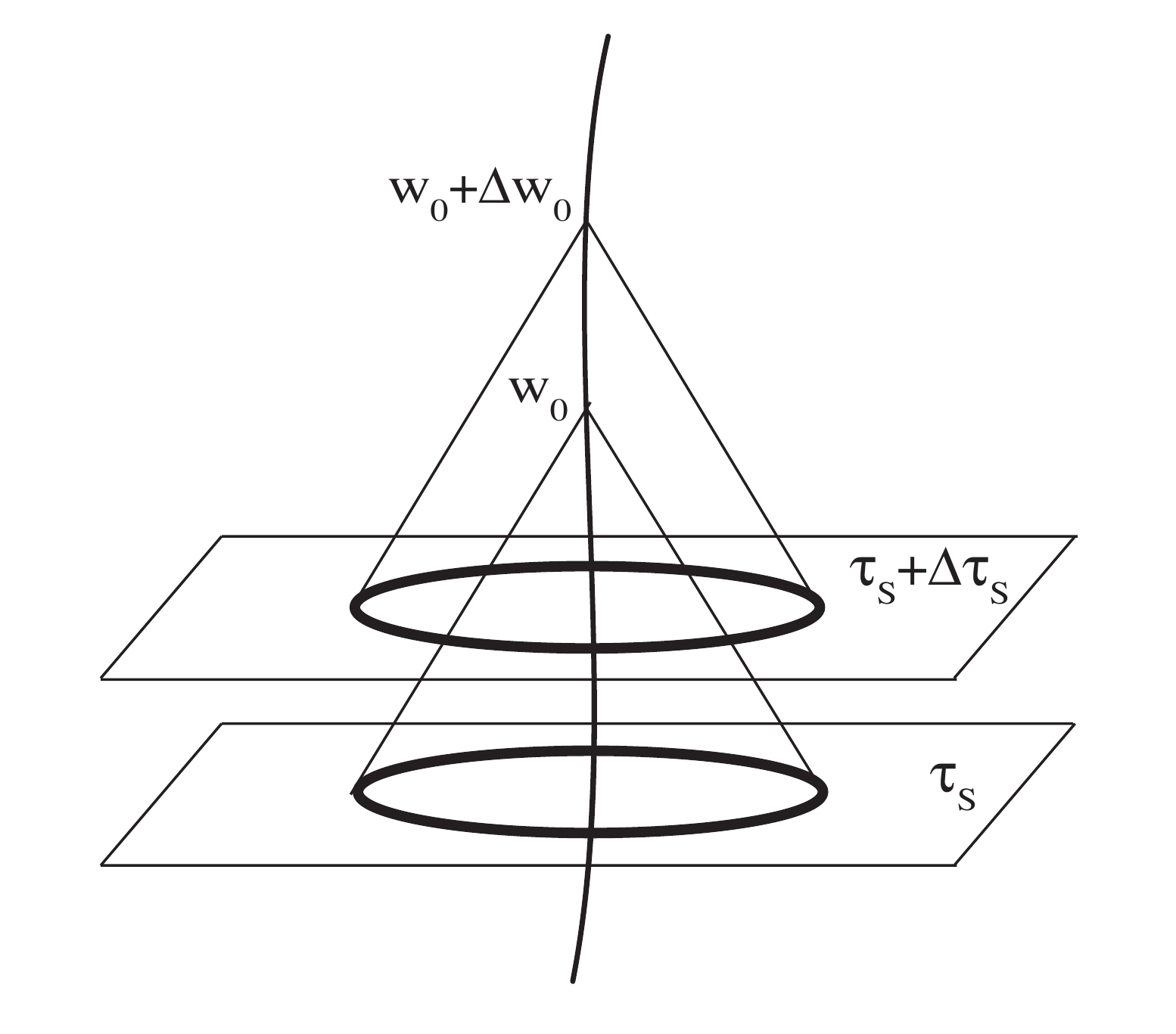}
 \caption{A graphic illustration of the redshift drift effect. A possible variation of the cosmological expansion rate is detected by comparing observations performed on two different past light-cones.}
 \label{fig2}
 \end{figure}

The redshift drift (RSD) is defined as the rate of change of the redshift of a given source with respect to the observer's proper time.
Since both the observer and the source simultaneously evolve in time, the relevant hypersurfaces for the RSD effect will consist of two light-cones with two different bases (see Fig. \ref{fig2}). Assuming that source and observer are in geodesic motion and have negligible peculiar velocity, the RSD in a FLRW Universe, and in the proper-time interval $\Da t_0$, can be simply expressed as (see e.g. \cite{QABCQ}):
\be
\frac{\Delta z}{\Delta t_0} = (1+z) H_0 - H_s = \frac{\dot{a}_0 - \dot{a}_s}{a_s},
\label{526}
\ee
 where $H_0$ is the present value of the Hubble parameter while  $H_s$ is its value at the emission time. Clearly the RSD effect is a good indicator of  cosmic acceleration as a function of $z$ which {\it does not} use hypothetical standard candles.

 In order to generalize this expression to a general inhomogeneous Universe (and attempt some averaging of it) 
let us consider again the geodesic observer of the GLC gauge, with coordinates $x^{\mu} = (w, \tau, \theta^a)$.
As shown in Eq. (\ref{zratioY}), $z$ is in principle  a function of seven independent variables,  namely  of $w_0$, $\tau_0$, $\theta^a_0$, 
$\tau_s$ and $\theta^a_s$ (note that, since
we are assuming that source and observer are on the same light-cone $w=w_0$ at $\tau=\tau_0$, they will be both on the  light-cone $w = w_0+\Delta w_0$ at $\tau= \tau_0+\Delta \tau_0$, see Fig. \ref{fig2}). 
We can also note that $\Ups_0$ is independent of $\theta^a_0$ (see the comment after Eq.(\ref{Int_z})).
As a consequence  we have only five independent contributions to the variation of $1+z$, and
 we can write, in general:
\be
 \Delta(1+z) = \frac{\partial (1+z) }{\partial w_0} \Delta w_0 + 
               \frac{\partial (1+z) }{\partial \tau_0} \Delta \tau_0 + 
 \frac{\partial (1+z) }{\partial \tau_s} \Delta \tau_s + 
 \frac{\partial (1+z) }{\partial \theta^a_s} \Delta \theta^a_s \; .
 \label{GVZ} 
 \ee

As shown in Eq.(\ref{FR}),  in the homogeneous limit the coordinate $\tau$ goes to the proper (cosmic) time $t$ of the synchronous gauge. So, locally around our geodesic observer, we choose to evaluate the redshift drift $\Da z$ with respect to his/her time parameter $\tau$,
and we need, to this purpose, an explicit relation between $\Da \tau_0$ and the variation of the other coordinates involved. 
For a geodesic observer with $n_\mu= -\pa_\mu \tau$ we have  $\dot{x}^{\mu} \sim  n^\mu=(1/\Upsilon, 1, U^a/\Upsilon)$, so that we can express $\Da x^\mu$ in terms of $\Da w$, at all times, as: 
\be
\Delta \tau=\Upsilon \Delta w \,\,\,\,\,\,\,\,\,\,\,\,\,,\,\,\,\,\,\,\,\,\,\,\,\,\,
\Delta \theta^a=U^a \Delta w .
\label{59}
\ee
 Using Eq. (\ref{59}) we find 
$\Da w_0=\Delta \tau_0/\Ups_0=\Delta \tau_s/\Ups_s$, 
and the final result for the RSD can be written in terms of $z$ and its derivatives as:
\be
\frac{ \Delta z}{\Delta \tau_0} = (1+z) \widetilde{H}_0 
+ \frac{1}{\Upsilon_0}{\pa \over \pa w_0} (1+z) 
+ {\pa \over \pa \tau_s} \ln(1+z) 
+ \frac{U^a_s}{\Upsilon_0}{\pa \over \pa \theta^a_s} (1+z) , 
\label{512}
\ee
where:
\be
\widetilde{H}_0\equiv \frac{1}{\Upsilon_0}\frac{\partial \Upsilon_0}{\partial \tau_0}
\label{Hetother}
\ee
(let us note again that $\partial_{w_0}$ acts on both metric factors $\Upsilon$ contained in $(1+z)$). This result is valid for a general inhomogeneous metric, and can be compared, as a useful consistency check, with a similar result previously obtained in the particular case of a spherically symmetric geometry \cite{Uzan}. If we move from our coordinates to the adapted coordinates used in \cite{Uzan} we find that our expression (\ref{512}) exactly reduces to the expression for the RSD  reported in Eq. (5) of Ref. \cite{Uzan}.

The quantities appearing in Eq. (\ref{512}) can now be averaged over the 2-sphere embedded, at $\tau=\tau_s$, in the light-cone $w=w_0$. Using our prescription based on Eq. (\ref{null2d_SLC_geodetic}) we find that both $\Ups_0$ and $\widetilde{H}_0$ 
can be taken out from the averaging integrals (which are performed over the variables $\theta^a$), and we obtain: 
\beq
{\l \Da z \re_{w_0, \tau_s} \over \Da \tau_0}= \l 1+z \re_{w_0, \tau_s} \widetilde{H}_0
+ \frac{1}{\Upsilon_0} \l \pa_{w} (1+z) \re_{w_0, \tau_s} 
+ \l \pa_{\tau} \ln(1+z) \re_{w_0, \tau_s} 
+ {1\over \Ups_0} \l U^a \pa_a (1+z) \re_{w_0, \tau_s} \,.
\eeq
For the derivatives performed with respect to the hypersurface parameters $w$ and $\tau$ we can now apply the Buchert-Ehlers commutation rules in the simplified form of Sect. \ref{Sec41}. Using in particular Eqs. (\ref{413}) and (\ref{414}) we are lead to:
\bea
{\l \Da z \re_{w_0, \tau_s} \over \Da \tau_0} &=& \l 1+z \re_{w_0, \tau_s} \widetilde{H}_0 
+ \frac{1}{\Upsilon_0} \pa_{w_0}  
\l 1+z \re_{w_0, \tau_s} 
+\pa_{\tau_s}  
\l \ln(1+z) \re_{w_0, \tau_s} 
\nonumber \\ & & 
+ Q_w(z)+Q_\tau(z),
\label{533}
\eea
where
\beq
Q_\tau(z)= - \left\langle \ln(1+z) \pa_\tau \ln\sqrt{|\ga|} \right\re_{w_0, \tau_s}
+\left\langle \ln (1+z) \right\re_{w_0, \tau_s} 
\left\langle  \pa_\tau \ln\sqrt{|\ga|} \right\re_{w_0, \tau_s},
\eeq
\bea
Q_w(z)&=&   \frac{1}{\Upsilon_0} \left\{-\left\l (1+z) \pa_w \ln\sqrt{|\ga|}  \right\re_{w_0, \tau_s} - \left\l (1+z)
\left[ \pa_a U^a+U^a \partial_a \ln\sqrt{|\gamma|} \right]  \right\re_{w_0, \tau_s}
\right.
\nonumber \\
& &\left. +\l 1+z  \re_{w_0, \tau_s} 
\left[\left\l \pa_w \ln\sqrt{|\ga|}  \right\re_{w_0, \tau_s} + \left\langle
 \pa_a U^a+U^a \partial_a \ln\sqrt{|\gamma|} \right\re_{w_0, \tau_s}\right]\right\}\,.
\eea
These last terms, together with the other terms of Eq. (\ref{533})  control the difference between the {\em averaged redshift drift} and the {\em drift} for the {\em averaged redshift} and represent  a light-cone analog of the backreaction terms due to the geometric inhomogeneities, first computed by Buchert \cite{1} in the context of three-dimensional spacelike averages. Unlike in that case, however, our backreaction is physically controlled by the geometric dynamics of a two-dimensional surface, with metric $\ga_{ab}$. Also, the possible physical meaning of these backreaction terms is in principle strongly model-dependent, and their physical effects are to be explicitly computed for any given model chosen to parametrize deviations from FLRW geometry.

For a homogeneous FLRW metric the averages disappear, and the backreaction terms $Q_\tau$ and $Q_w$ are identically vanishing. In that case $\widetilde{H}_0=H_0$ and, using the homogeneous limit of 
Eq.(\ref{FR}), one also finds $\pa_w(1+z)=0$ and $\pa_\tau \ln(1+z)_s=-H_s$, so that Eq. (\ref{533})  gives back the result (\ref{526}). For a general inhomogeneous model, however, several new contributions appear. It would be very interesting to see whether such backreaction effects are relevant for determining the 
kinematics of the Universe (and the equation of state of its  energy components as a function of $z$) from the forthcoming RSD experiments (see e.g. \cite{CODEX}). We plan to come back to this question elsewhere \cite{BGMNV}, at least within specific inhomogeneous models such as the Lema\^itre-Tolman-Bondi (LTB) Universe \cite{LTB}.


\section{Conclusions}
\label{Sec6}
\setcounter{equation}{0}

In this paper we have made a first step towards defining suitable covariant and gauge invariant light-cone  averages that should be  relevant to analyze the effects of inhomogeneities on astrophysical observables related to light-like (massless) signals.
We were able to define such averages both on light-like hypersurfaces and on ordinary 2-surfaces embedded on a specific  light-cone (e.g. our past one).
We have also written down  the generalized version of the Buchert-Ehlers commutation rules (between averaging and differentiation) that hold for all such light-cone averages.

One obvious problem to which we would like to apply our technique is that of cosmic acceleration. This is actually nothing but a measured relation between redshift and luminosity distance  which can only be explained in a FLRW cosmology if  the Universe underwent a late-time accelerated expansion. Within General Relativity it implies, in turn, the existence of a large dark energy component in the cosmic fluid. Our formalism allows in principle to study directly the luminosity-redshift relation for a given non homogeneous Universe (e.g of the LTB type \cite{LTB}) or, even more interestingly, for  a Universe containing a stochastic spectrum of inhomogeneities like those that originate from inflationary cosmology.

In order to prepare the ground for such investigations 
we have introduced some adapted  ``geodesic null" coordinates  that allow to express in a simple way redshift and luminosity distance in terms of local metric components (i.e. calculated at the source's position). Furthermore, our expressions for both  the averages and for their derivatives take a considerably simpler form in these coordinates.

We have also applied our formalism to the case of the so-called redshift drift, a quantity which, without any need of using standard candles, should be sensitive to a possible acceleration of the scale factor at different redshift values, and thus could distinguish among different forms of dark energy. Once more, however, one should address the question of how inhomogeneities could modify the FLRW relation between redshift drift and expansion rate.
By using the geodesic light-cone gauge we were able to find the formal modification of the relation and show that it contains, among others, Buchert-Ehlers-like commutator terms. The physical interpretation of such terms is, however, model-dependent, and their possible effect are to be extracted for any given model with an explicit computation.

Unlike in the case of spacelike hypersurfaces we have found that the most useful averages for physical applications are related to surface averages. However, it is easy to see that the quantities to be averaged are themselves sensitive to inhomogeneities over the whole past light-cone hypersurface even when they can be expressed in terms of quantities living on the surface. So far we have not found physical examples in which a true average over the full light-cone is involved, and we strongly suspect that there isn't one.

In order to find out whether the formalism we have introduced is actually useful one will have to perform detailed calculations within realistic and   explicit models for the relevant inhomogeneities. These can be either abstract mathematical ones, like LTB, swiss-cheese, or fractal models, or realistic ones based on our present knowledge of the large scale structure. 

Another interesting exercise would be to compute the magnitude of several effects to second order in a perturbed FLRW Universe. 
Even if, as most people suspect, the effect of inhomogeneities will turn out to be small and not to change the conclusions that follow from simple FLRW consideration,  the result of such an investigation will be important in that it will sharpen considerably  the conclusion that dark energy is indeed unavoidable.

 
\section*{Acknowledgements}

We wish to thank Ido Ben-Dayan for useful discussions during the early stages of this investigation. Useful correspondence with Luca Amendola, Roy Maartens and Misao Sasaki is also gratefully acknowledged.


\end{document}